\documentclass[nolinenumbers, RNAAS]{aastex631}
\usepackage{graphicx} 
\usepackage[utf8]{inputenc}
\usepackage{amssymb}
\usepackage{natbib}

\defcitealias{Behroozi:2019}{B19}
\defcitealias{kaushal:23}{K23}

\begin{document}

\title{A Comparison of Star-Formation Histories Derived from UniverseMachine \\ and LEGA-C at 0.6 $<$ z $<$ 1}

\author[0009-0009-9245-4651]{Cecilia Steel}\affiliation{Department of Physics \& Astronomy and PITT PACC, University of Pittsburgh, Pittsburgh, PA 15260, USA}

\author[0000-0001-9820-9619]{Alan Pearl}\affiliation{HEP Division, Argonne National Laboratory, 9700 South Cass Avenue, Lemont, IL 60439, USA}
\affiliation{Department of Physics \& Astronomy and PITT PACC, University of Pittsburgh, Pittsburgh, PA 15260, USA}

\author[0000-0003-4382-4467]{Yasha Kaushal}\affiliation{Department of Physics \& Astronomy and PITT PACC, University of Pittsburgh, Pittsburgh, PA 15260, USA}

\author[0000-0001-5063-8254]{Rachel Bezanson}\affiliation{Department of Physics \& Astronomy and PITT PACC, University of Pittsburgh, Pittsburgh, PA 15260, USA}


\begin{abstract}

 In this work, we compare star formation histories of massive (10.5 $< \log(\mathrm{M_*/M_{\odot}}) <$ 12) galaxies in the UniverseMachine model to those measured from the Large Early Galaxy Astrophysics Census (LEGA-C) at $0.6<z<1$. Following the LEGA-C study, we investigate how 50\% ($t_{50}$) and 90\% ($t_{90}$) formation timescales depend on total stellar mass. We find good agreement between the observed and model timescales for the star-forming population $\Delta\,t_{SF}\lesssim1\,\mathrm{Gyr}$ across the full mass range. In contrast, the observed age-mass correlation is weaker for the quiescent population compared to UniverseMachine models ($\Delta t_{Q}\lesssim2\,\mathrm{Gyr}$), especially at the high-mass end. This indicates continued star formation or additional processes in the most massive quiescent galaxies, a behavior not accounted for in the UniverseMachine model. 




\end{abstract}

\section{Introduction}
The UniverseMachine (\citet{Behroozi:2019}, hereafter \citetalias{Behroozi:2019}) is a tool that generates empirically-constrained galaxy demographics across cosmic time, including growth histories, by correlating individual star formation histories (SFHs) with halo accretion rates. This model is physically motivated by simulations and is tuned to accurately reproduce a wide variety of observed galaxy populations. The empirical relations included have been derived from various observational studies of the average star formation rates (SFRs) of star-forming galaxies, quenched fractions, and more, spanning a wide range of redshifts \citepalias[][and references within]{Behroozi:2019}.

The UniverseMachine is only calibrated to reproduce instantaneous SFRs, not ages or higher-order moments of SFHs. This presents an avenue to test and improve the model. Previous studies have compared these SFHs of local universe galaxies with the ones derived from their broadband photometric data (\citetalias{Behroozi:2019}, \citet{Leja_2019}). \citetalias{Behroozi:2019} showed UniverseMachine agrees best with observational stellar growth rates \citep{Pacifici:2016} at early times, but the massive galaxies grow too rapidly in the empirical models. \citet{Leja_2019} and \citet{Carnall:2019} 
found that parametric models systematically underestimate late formation times while non-parametric models overestimate early formation times by 50\% and 25\% respectively with respect to the UniverseMachine. 
No studies have yet investigated the nature of SFHs at higher redshifts derived empirically and observationally owing to the dearth of high-quality statistical spectroscopic data. \par
 
The Large Early Galaxy Astrophysics Census (LEGA-C) \citep{van_der_Wel_2021,Straatman_2018,vanderwel:16} provides a high-S/N mass-complete spectroscopic sample of $\sim{3000}$ massive galaxies ($\log(M_{\star}/M_{\odot})\gtrsim10.5$) at half the age of the Universe. 
\citet{kaushal:23} (hereafter \citetalias{kaushal:23}) measured SFHs of the full LEGA-C sample and quantified their population trends with stellar mass and stellar velocity dispersion. This analysis is especially powerful because even massive galaxies in the LEGA-C sample are younger than their local counterparts. In contrast, their descendants today are old, rendering precise age determination notoriously challenging. In this study, we construct a mass-and-redshift matched sample of UniverseMachine model galaxies and compare the SFH population trends with those published in \citetalias{kaushal:23}.


\section{Data and Methods}

\subsection{Stellar population synthesis (SPS) modeling of LEGA-C}
\label{sec:legac-methods}
This work leverages \citetalias{kaushal:23} SFHs derived from SPS modeling of the LEGA-C dataset. We refer the reader to that paper for full details of the analysis. The measurements were made using two Bayesian SPS tools: \texttt{Bagpipes}  \citep{carnall:18} (parametric) and \texttt{Prospector} \citep{johnson:17,Johnson:21} (non-parametric). Each adopts a range of different modeling priors, most notably the treatment of SFHs as either analytical functions or as more flexible piecewise constant functions.
We adopt $t_{50}$ and $t_{90}$ from \citetalias{kaushal:23}, 
corresponding to the times a galaxy forms 50\% and 90\% of its total stellar mass, uncorrected for mass-loss. 
In this work we use \citetalias{kaushal:23} trends in $t_{50}$ and $t_{90}$ with stellar mass as our empirical benchmark. 

\subsection{SFHs from the UniverseMachine}
The UniverseMachine imposes a physically motivated correlation between the assembly of galaxies and their dark matter halos, producing SFHs that may be more tied to physical growth of a galaxy than the output of SPS modeling. We employ the UniverseMachine Data Release 1 results derived from the Bolshoi-Planck simulation. We sample UniverseMachine galaxies within 1.5$\times10^{7}h^{-3}Mpc^3$ such that they follow the LEGA-C sample redshift distribution from 13 evenly spaced snapshots between $0.6<z<1$. We construct SFHs for each model galaxy by connecting the SFRs of the progenitors of these galaxies at all previous snapshots.
We make the simplifying approximation of following only the most massive progenitors and verify that this approach is robust by testing against full-merger-tree SFHs from \citetalias{Behroozi:2019} at z=0.7. 
We find very good agreement for $t_{50}$, and up to $\sim{10\%}$ discrepancies in $t_{90}$, which is well below our level of uncertainty from SPS modeling.
\section{Results}

\begin{figure}[ht!]
\plotone{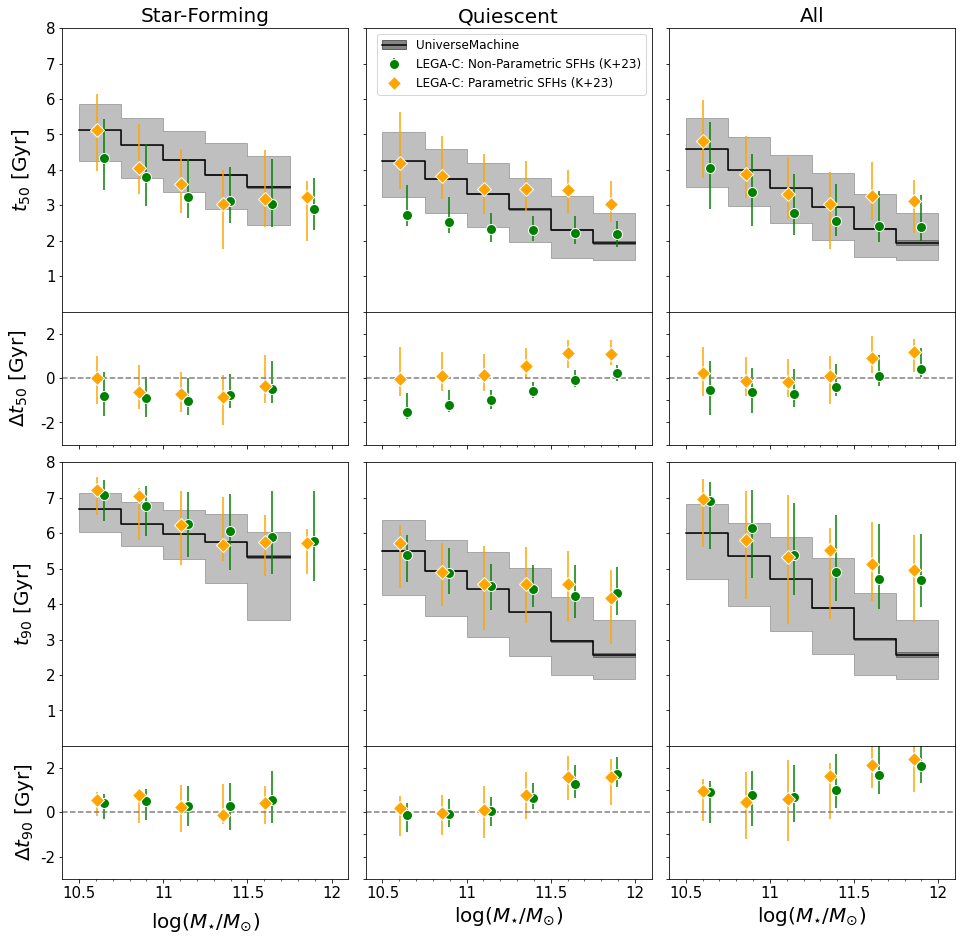}
\caption{$t_{50}$ and $t_{90}$ formation times (forward in time) for star-forming (left), quiescent (middle), and all galaxies (right). Black lines represent UniverseMachine medians with dark gray errors and light gray 16-84 percentile population scatter. Colored points depict LEGA-C population trends (median and population scatter). Residuals between the UniverseMachine and LEGA-C medians are shown below. 
\label{fig:1}}
\end{figure}

Figure \ref{fig:1} shows the population trends of $t_{50}$ and $t_{90}$ with stellar mass. 
Colored symbols show LEGA-C measurements, while gray bands represent UniverseMachine trends. UniverseMachine errors encompass Poisson error on medians and systematic offsets from main-progenitor-only search. In all cases, the data and models reproduce the well-established trend that more massive galaxies host older stars. 

This comparison consistently suggests that star-forming galaxies are $\sim1\,\mathrm{Gyr}$ older than predicted by the UniverseMachine. Similarly, we find disagreement in the onset and duration of early star-formation amongst the quiescent population. Although UniverseMachine trends are in better agreement with \texttt{Bagpipes}, the empirical discrepancy highlights that this measurement is strongly influenced by modeling priors. 
Generally, late formation times agree for star-forming galaxies at all masses and for low-mass quiescent populations ($\Delta t_{90}\lesssim1\,\mathrm{Gyr}$). In contrast, the most massive quiescent galaxies are formed ${\sim}1-2\,\mathrm{Gyr}$ earlier in UniverseMachine than suggested by observations. 
We find that UniverseMachine predictions for star-forming galaxies agree well with observations. However, observed SFHs suggest more extended and gradual growth. 
For quiescent galaxies, the observed age-mass correlation is much stronger in the UniverseMachine than in the observations, especially at the high-mass end. This points to scope of improvement in robustly recovering early stellar mass growth of quiescent galaxies from SPS modeling as well as late time assembly of massive galaxies in the UniverseMachine at this epoch.

\section{Acknowledgements}

This research was supported in part by the University of Pittsburgh Center for Research Computing using the H2P cluster, which is supported by NSF Award \#OAC-2117681. RB, YK, and CS gratefully acknowledge support from NSF-CAREER grant AST-2144314.

\bibliography{bib}
\end{document}